# Enhancing the impedance matched bandwidth of bottle microresonator signal processing devices

**M. Sumetsky and S. Zaki**

*Aston Institute of Photonic Technologies, Aston University, Birmingham B4 7ET, UK*
*Corresponding authors: m.sumetsky@aston.ac.uk; zakis@aston.ac.uk*

**Light pulses entering an elongated bottle microresonator (BMR) from a transversely oriented input-output waveguide (microfiber) slowly propagate along the BMR length and bounce between turning points at its constricting edges. To avoid insertion losses and processing errors, a pulse should completely transfer from the waveguide into the BMR and, after being processed, completely return back into the waveguide. For this purpose, the waveguide and BMR should be impedance matched along the pulse bandwidth. Here we show how to enhance the impedance matched bandwidth by optimization of the BMR effective radius variation in a small vicinity of the input-output waveguide.**

Engineering of microphotonic signal processing devices with the smallest possible dimensions is often based on the slow light concept [1]. In special cases (e.g., fabrication of miniature broadband delay lines), miniaturization can be achieved by bending of photonic waveguides in 2D or 3D without interturn coupling, e.g., into a spiral [2, 3] or coil [4, 5] (Fig. 1(a) and (b)). Advanced signal processing and greater miniaturization is achieved in slow light devices created using photonic crystals [6, 7], ring resonators [8, 9] (Fig. 1(c) and (d)), and coiled waveguides (Fig. 1(b)) with interturn coupling [10]. In these devices, the decrease in the effective propagation speed of light is achieved by multiple reflections and circulations, which results in dramatic dispersion and reduction of transmission bandwidth. One of the significant problems on the way to improve the performance of slow light devices is impedance matching (IM), i.e., in suppressing losses and reflections in the regions of their coupling with the input and output waveguides. To solve this problem (which is naturally absent in in the case of bent waveguide devices illustrated in Fig. 1(a) and (b)), slow light devices are partitioned into IM regions and signal processing regions, as illustrated in Fig. 1 (c) for photonic crystals [11, 12] and in Fig.1 (d) for coupled ring resonators [13]. These regions can be optimized separately.

Several years ago, a conceptually different low loss slow light signal processing device based on the slow whispering gallery mode (WGM) propagation along the elongated bottle microresonator (BMR) created at the surface of an optical fiber by nanoscale effective radius variation (ERV) illustrated in Fig. 1(e) was proposed and experimentally demonstrated [14]. The ERV of this BMR was introduced with subangstrom precision using the surface nanoscale axial photonics (SNAP) technology [15]. The fabricated SNAP BMR had the ERV shape of a 2.8 nm high and 3 mm long semi-parabola. It exhibited the 2.58 ns delay of 100 ps pulses with less than 0.5 dB/ns loss and without dispersion. It was demonstrated in [14] that it is possible to achieve the IM condition near the selected wavelength by tuning the coupling with the input-output microfiber and its axial position.

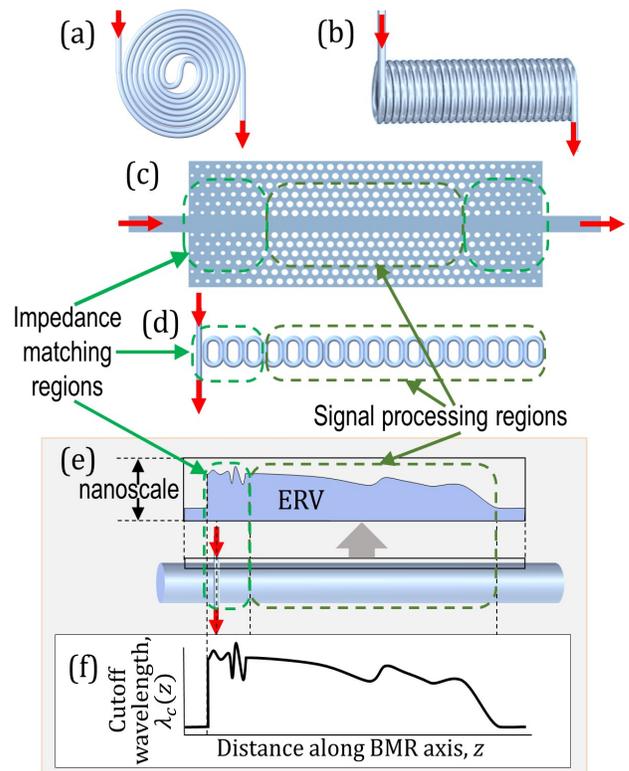

Fig. 1. (a) Spiral delay line. (b) Coil delay line. (c) Photonic crystal connected to the input and output waveguide. (d) Couple ring microresonators connected to the input-output waveguide. (e) SNAP BMR with nanoscale ERV (shown in the inset) connected to the input-output waveguide (microfiber). (f) Cutoff wavelength variation corresponding to the ERV in (e).

Here we show that the IM bandwidth of a BMR can be significantly enhanced by optimization of its ERV in a small vicinity of the transverse input-output microfiber. In our calculations, we follow the theory of Ref [15]. The wavelengths $\lambda$ of WGMs launched into the BMR from the transverse microfiber must be close to the BMR cutoff wavelength $\lambda_c(z)$ (Fig. 1(f)) to ensure slow propagation of light along the BMR axis $z$.

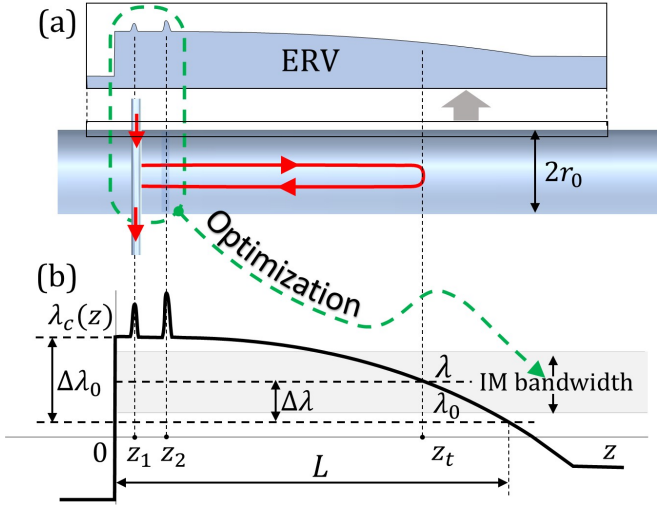

Fig. 2. SNAP BMR with semi-parabolic ERV (a) and corresponding cutoff wavelength variation (b). The IM matching region circled in (a) includes the microfiber position $z_1$ and position $z_2$ of the local ERV modification which is optimized to enhance the IM bandwidth.

To demonstrate our idea, we consider a slow light dispersionless delay line BMR of length $L$, which *prior to optimization* has the semi-parabolic shape of the cutoff wavelength variation [14] (Fig. 2),

$$\Delta\lambda_c(z) = \begin{cases} \Delta\lambda_0(1 - z^2/L^2) & z > 0 \\ -\infty & z < 0 \end{cases} \quad (1)$$

The sharp cut at $z = 0$ can be introduced, e.g., with a femtosecond laser [16]. Here, for convenience, we introduce the bandwidth of our concern $\Delta\lambda_0$ starting from wavelength $\lambda = \lambda_0$ so that the cutoff wavelength variation $\Delta\lambda_c(z) = \lambda_c(z) - \lambda_0$ and the wavelength variation $\Delta\lambda = \lambda - \lambda_0$. Then variation of the WGM amplitude along the axis $z$ satisfies the one-dimensional wave equation (see e.g. [15, 17]):

$$\frac{d^2\Psi(z)}{dz^2} + \beta^2(\lambda, z)\Psi(z) = 0,$$
$$\beta(\lambda, z) = 2^{3/2}\pi n(\lambda_c)^{-3/2}(\Delta\lambda_c(z) + i\gamma - \Delta\lambda)^{1/2}. \quad (2)$$

Here $n$ and $\gamma$ are the refractive index and propagation loss of the fiber material and $\beta(\lambda, z)$ is the WGM propagation constant. The ERV corresponding to $\Delta\lambda_c(z)$ is found by its rescaling as $\Delta r_{eff}(z) = r_0 \Delta\lambda_c(z)/\lambda_0$ where $r_0$ is the BMR radius. The amplitude $S(\lambda, z_1)$ through the input-output microfiber positioned at $z = z_1$ (Fig. 2) is determined as [15]

$$S(\lambda, z_1) = \frac{1 + D_1^* G(\lambda, z_1, z_1)}{1 + D_1 G(\lambda, z_1, z_1)}. \quad (3)$$

Here complex parameter $D_1$ determines the effect of the microfiber on the WGM propagation along the BMR and $G(\lambda, z_1, z_1)$ is the Green's function of Eq. (2). Eq. (3) was obtained under the assumption of lossless coupling [18] when the effect of the microfiber is determined by substitution $\beta^2(\lambda, z) \to \beta^2(\lambda, z) + D_1\delta(z - z_1)$ and addition of source $2\text{Im}(D_1)\delta(z - z_1)$ in Eq. (2) [15]. Modelling of the microfiber by the delta function $\delta(z - z_1)$ is justified since its cross-sectional dimension (~ 1 µm) is much smaller that the characteristic slow WGM axial wavelength (~100 µm).

The group delay is determined from Eq. (3) as

$$\tau(\lambda, z_1) = -\frac{\lambda_0^2}{2\pi c}\text{Im}\left(\frac{\partial \ln(S(\lambda, z_1))}{\partial \lambda}\right), \quad (4)$$

where $c$ is the speed of light. Assuming that the length of the of BMR is sufficiently long (to ensure nanosecond scale delay times) we solve Eq. (2) over the signal processing region in the semiclassical approximation. First, we assume that the ERV in the much shorter IM region near $z = 0$ is not optimized so that we can set there $\beta(\lambda, z) = \beta(\lambda, 0)$. In this region, we define two solutions of Eq. (2), $\Psi_1(z)$ and $\Psi_2(z)$ satisfying the boundary conditions $\Psi_1(0) = 0$ and $\Psi_2(z) \to 0$ away from the turning point, $z > z_t$ (Fig. 2(b)). The Green's function in the IM region is expressed through these solutions as $G_0(\lambda, z_1, z_2) = \Psi_1(z_<)\Psi_2(z_>)/W(\Psi_1, \Psi_2)$ where $W$ is their Wronskian, $z_> = \max(z_1, z_2)$, $z_< = \min(z_1, z_2)$, and

$$\Psi_1(z) = \sin(\beta(\lambda, 0)z), \quad \Psi_2(z) = \cos(\varphi(\lambda) - \beta(\lambda, 0)z),$$
$$\varphi(\lambda) = 2^{1/2}\pi^2 nL\lambda_0^{-3/2}(\Delta\lambda_0)^{-1/2}(\Delta\lambda_0 + i\gamma - \Delta\lambda) + \frac{\pi}{4}. \quad (5)$$

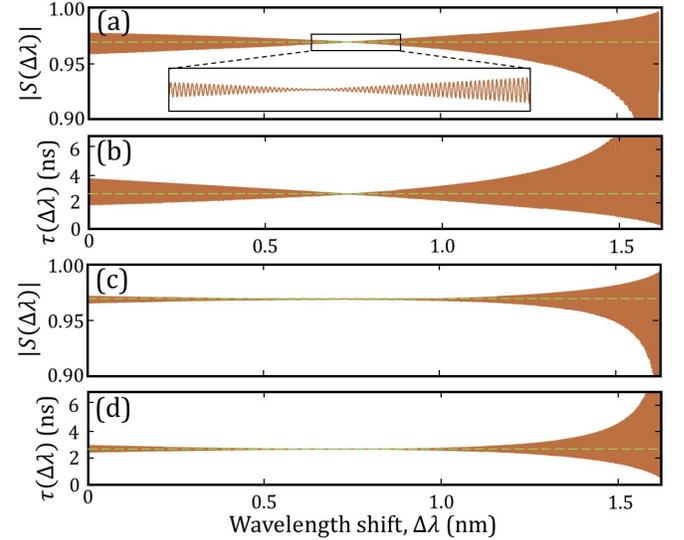

Fig. 3. (a) Transmission amplitude $|S(\lambda, z_1)|$ and (b) group delay $\tau(\lambda, z_1)$ for the BMR with the IM bandwidth optimized by variation of coupling parameter $D_1$ and microfiber position $z_1$. Inset in (a) magnifies oscillations of the transmission amplitude, which are similar in all plots presented in this figure. (c) and (d) are the same as in (a) and (b) but now for the BMR with IM bandwidth optimized by variation of microfiber parameters $D_1$, $z_1$ and ERV parameters $D_2$ and $z_{12}$. Dashed green lines in all plots are the averaged spectra.

The transmission amplitude $S(\lambda, z_1)$ with this Green's function is impedance matched at wavelength $\lambda = \lambda_{IM}$ if the coupling parameter $D_1$ and microfiber position $z_1$ satisfy the equation

$$D_1 = \beta(\lambda_{IM},0)\big(\cot(\beta(\lambda_{IM},0)z_1)+i\big), \qquad (6)$$

coinciding with the known result [14]. In the classical approximation, the group delay found from Eqs. (4)–(6) is $\tau(\lambda,z_1) = \tau_{cl} = 2^{-1/2}\pi n L(\lambda_0/\Delta\lambda_0)^{1/2}/c$ [14]. It follows from Eq. (6) that, since the larger IM bandwidth assumes larger value of propagation constant $\beta(\lambda_{IM},0)$, it also requires a larger imaginary part of the coupling parameter determined from Eq. (6) as $\mathrm{Im}(D_1) = \beta(\lambda_{IM},0)$. At the same time, smaller $z_1$ leads to slower variation of $\mathrm{Re}(D_1)$ as a function of $\beta$ and, thus, to wider IM bandwidth.

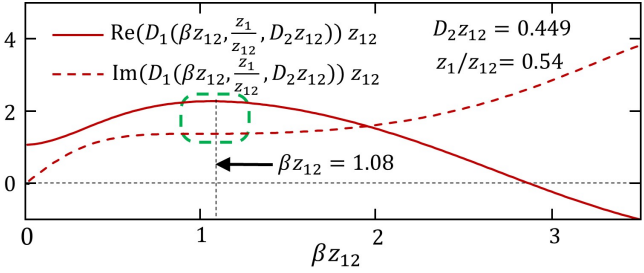

Fig. 4. Optimization of the BMR dimensionless parameters $\xi = \beta(\lambda_{IM},0)z_{12}, \rho = z_1/z_{12}$, and $\Xi_j = D_j z_{12}$.

In Figs. 3(a) and (b), we show the behavior of the transmission amplitude, $|S(\lambda,z_1)|$, and group delay, $\tau(\lambda,z_1)$ at wavelength $\lambda_0 = 1550$ nm and BMR Q-factor $Q = 10^8$ [19, 20], which corresponds to $\gamma = \lambda_0/Q = 0.155$ pm. To arrive at the group delay $\tau_{cl} = 2.6$ ns, similar to that experimentally observed in Ref. [14], but with much larger IM bandwidth of 1 nm, we followed the rescaling relations [14] setting the BMR length $L = 7.9$ mm and the bandwidth introduced in Eq. (1) $\Delta\lambda_0 = 1.6$ nm. We chose the position of exact IM wavelength at $\lambda_{IM} = \lambda_0 + \Delta\lambda_{IM}$ with $\Delta\lambda_{IM} = 0.73$ nm, where the propagation constant is $\beta(\lambda_{IM},0) = \mathrm{Im}(D_1) = 0.2\ \mu\mathrm{m}^{-1}$, and set $\mathrm{Re}(D_1) = \mathrm{Im}(D_1)$ [21]. The calculated microfiber position for these parameters was $z_1 = 3.9$ μm. The IM condition results in minimized oscillations of the BMR transmission amplitude and group delay spectra presented in Figs. 3(a) and (b) which indicates suppression of reflections from the coupling region.

We tested the performance of the designed delay line by modeling the propagation of a 100 ps pulse. In the case of perfect IM bandwidth, the pulse should be completely transmitted back into the input-output microfiber without dispersion after a single round trip in the BMR having acquired a delay time of $\tau_{cl} = 2.6$ ns. The deviation from the IM condition generates a cascade of pulses due to successive reflections, the strongest of which directly propagates into the output part of the microfiber without even entering the BMR (Fig. 5(a)). Blue curve in Fig. 5(b) shows the dependence of the ratio of amplitude $A_s$ of this undelayed pulse and the amplitude $A_0$ of the input pulse in the 1 nm wavelength interval including the IM wavelength $\lambda_{IM}$. As follows from Eq. (6), the widest IM bandwidth surrounding $\lambda_{IM}$ corresponds to the smallest microfiber coordinate and largest possible $\mathrm{Re}(D_1) = z_1^{-1}$, which may not be practically feasible. However, the improvement of the IM bandwidth in the limit $z_1 \to 0$ is not significant (dashed blue curve in Fig. 5(b)).

Now we show that optimization of the cutoff wavelength profile $\Delta\lambda_c(z)$ in a small vicinity of the microfiber position can significantly enhance the IM bandwidth. In the example considered, we modify $\Delta\lambda_c(z)$ defined by Eq. (1) locally, near axial position $z_2$, by adding $\Lambda_2\delta(z-z_2)$ with real $\Lambda_2$. In practice, this modification can be introduced, e.g., by a femtosecond laser [16, 22], local coating, or an adiabatic phase-unmatched contact with a microfiber to avoid losses and azimuthal reflections. The transmission amplitude in Eq. (3), is now expressed through the modified Green's function

$$G(\lambda,z_1,z_1) = G_0(\lambda,z_1,z_1) - \frac{D_2 G_0^2(\lambda,z_1,z_2)}{1+D_2 G_0(\lambda,z_2,z_2)},$$

$$D_2 = 8\pi^2 n^2 (\lambda_c)^{-3}\Lambda_2. \qquad (7)$$

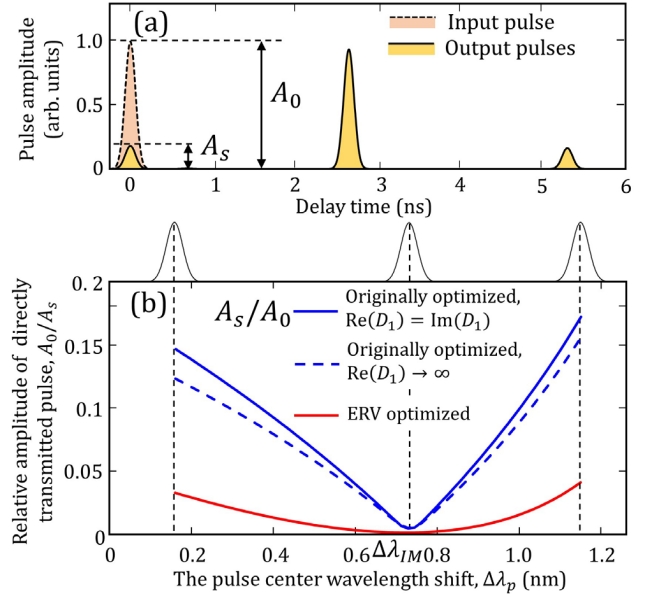

Fig. 5. (a) The 100 ps input pulse (dashed orange) and the train of output pulses (solid yellow). For ideal IM device, there is only a single output pulse at time 2.6 ns. The deviation from the IM condition gives rise to undelayed and multiply reflected pulses outgoing from BMR at times 0, 5.2, 7.8, … ns. (b) Ratio of the amplitude of the pulse outgoing without delay and the amplitude of the input pulse $A_s/A_0$ as a function of the pulse central wavelength shift. Solid blue curve corresponds to the BMR with spectra shown in Fig. 3(a) and (b) optimized for $\mathrm{Re}(D_1) = \mathrm{Im}(D_1)$. Dashed blue curve corresponds to $\mathrm{Re}(D_1) \to \infty$ and solid red curve corresponds to the BMR with microfiber and ERV parameters optimized as shown in Fig. 4. The pulses at the top of (b) are spectra of a 100 ps pulse with central wavelengths at the beginning and end of the 1 nm wavelength band and at $\lambda_0 + \Delta\lambda_{IM}$.

The IM condition, which we derived from Eqs. (3), (5), and (7), is convenient to express through the dimensionless parameters in the form:

$$\Xi_1 = \xi\left(\cot(\rho\xi) + \frac{i\xi\exp(i\xi) - \Xi_2\cos(\xi)}{\xi\exp(i\xi) - \Xi_2\sin(\xi)}\right), \qquad (8)$$

$$\xi = \beta(\lambda_{IM},0)z_{12}, \quad \rho = z_1/z_{12}, \quad \Xi_j = D_j z_{12},$$

where $z_{12} = z_2 - z_1$. For the original BMR shape defined by Eq. (1), i.e., for $\Xi_2 = 0$, this equation coincides with Eq. (5). In order to ensure the approximate fulfillment of Eq. (8) at maximum possible bandwidth, we minimize the derivative $d\Xi_1/d\xi$ found from Eq. (8) in the neighborhood of the same wavelength $\lambda_{IM} = \lambda_0 + \Delta\lambda_{IM}$ with $\Delta\lambda_{IM} = 0.73$ nm and $\beta(\lambda_{IM},0) = 0.2\ \mu\mathrm{m}^{-1}$, where the IM condition is satisfied exactly. The result of minimization shown in Fig. 4 yields $\xi = \beta(\lambda_{IM}, 0)z_{12} = 1.08$, $\rho = z_1/z_{12} = 0.54$, and $\Xi_j = D_j z_{12} = 0.449$. Thus, for $\beta(\lambda_{IM}, 0) = 0.2$ μm$^{-1}$, we have $z_1 = 2.64$ μm, $D_1 = 0.429 + 0.261i$ μm$^{-1}$, and $D_2 = 0.092$ μm$^{-1}$. The transmission amplitude and group delay with these parameters are shown in Fig. 3(c) and (d). Significant reduction of oscillations in the group delay and transmission amplitude spectrum compared to that shown in Fig. 3 (a) and (b) is seen. Fig. 5(b) compares the ratios of the amplitudes of the input and undelayed pulses, $A_s/A_0$, for the case of originally optimized device (blue curves) and the case with the ERV optimized (red curve). This figure confirms significant reduction of the undelayed pulse amplitude in the latter case. No noticeable pulse dispersion was observed in both cases.

For the BMR Q-factor $Q = 10^8$ and delay time 2.6 ns considered, the attenuation of pulses due to material losses are small (Fig. 3(a) and (c).) It is therefore interesting to determine the maximum possible delay that can be exhibited by our device without significant losses. The pulse amplitude $A_r$ after the roundtrip inside the BMR is simply expressed through the complex part of phase $\varphi(\lambda)$ in Eq. (4) and classical delay time $\tau_{cl}$ as $A_r = \Theta A_0$ where

$$\Theta = \exp\left(-\frac{2^{1/2}\pi^2 nL}{Q(\lambda_0 \Delta\lambda_0)^{1/2}}\right) = \exp\left(-\frac{2\pi c \tau_{cl}}{Q\lambda_0}\right). \quad (9)$$

From this equation, which accurately coincides with the numerical value of average transmission amplitude found from Figs. 3(a) and (c), we find that at wavelength $\lambda_0 = 1550$ nm a SNAP BMR with $Q = 10^8$ can perform the delay of optical pulses as long as 50 ns within the 1 nm bandwidth with $\sim 50\%$ attenuation. The length of this BMR is around 15 cm for 1 nm bandwidth and 5 cm for 0.1 nm bandwidth.

The SNAP BMR delay line has a cylindrical geometry similar to that of a coil delay line (Fig. 1(b)). While a low loss microscopic coil delay line has not yet been demonstrated [4, 5], it is interesting to compare its performance with the BMR delay line considered here. To avoid the interturn coupling, we set the coil pitch equal to $2\lambda_0$. Then, the surface of a cylinder with length $L$ and radius $r_0$ can fit a coil of length $L_{coil} = \pi r_0 L/\lambda_0$. If fabricated from the same (silica) material, the coil exhibits the delay time $\tau_{coil} = nL_{coil}/c$ and amplitude attenuation $\Theta_{coil} = \exp(-4\pi^2 \gamma n r_0 \lambda_0^{-3} L)$. The ratio of delay times $\tau_{cl}/\tau_{coil} = 2^{1/2}\lambda_0^{3/2}(\Delta\lambda_0)^{-1/2}r_0^{-1}$ of BMR and coil with the same footprint $r_0 L$ is around 1.8 for the parameters we used above and can be greater for smaller bandwidth and BMR radius. The attenuations of such coil and BMR satisfy the relation $\ln(\Theta)/\ln(\Theta_{coil}) = \frac{1}{2}\tau_{cl}/\tau_{coil} \cong 1$, i.e., approximately the same.

In conclusion, we showed that the IM bandwidth of a SNAP BMR signal processing device (in particular, a miniature delay line) can be enhanced by modification of the ERV or, equivalently, the cutoff wavelength variation of the optical fiber. A similar approach can be applied to other BMR-based devices (e.g., dispersion compensators [23] and advanced nonlinear SNAP devices [24]). The delta-function model used above for the ERV optimization is feasible and may be realized, e.g., with a contact to an optical microfiber. Generally, it is of great interest to determine feasible ERV profiles which ensure the IM matching condition with the best precision over the largest possible bandwidth. We note that increasing the IM bandwidth can also be achieved by rescaling the height and length of the BMR [14]. However, the direct application of the latter approach to larger IM transmission bandwidths requires impractically small separation between the input-output microfiber and the cut end of the BMR. All calculations above were performed in the vicinity of a cutoff wavelength which is usually specified by radial and azimuthal quantum numbers, $p$ and $m$ respectively [17]. Due to the quasi-periodicity of the transmission amplitude dependence on $m$ for the WGMs with $m \gg 1$ considered, it is accurately reproduced near several adjacent cutoff wavelengths with different $m$ and the same $p$. Therefore, we suggest that a much broader bandwidth multichannel device can be created as well. To this end, the WGMs with radial quantum numbers except for the fundamental one with $p = 0$ have to be suppressed using, e.g., a microcapillary fiber with a thin wall (see e.g., [25] and references therein) and the fiber radius should be adjusted to arrive at the required azimuthal free spectral range.

**Funding.** Engineering and Physical Sciences Research Council, (EP/P006183/1).

**Disclosures**. The authors declare no conflicts of interest.